\documentclass[letter]{jpconf}
\usepackage{graphicx}
\begin{document}
\title{Using observations of millisecond pulsars to measure mass and radius of neutron stars and implications for equation of state of matter at high density}

\author{Denis Leahy}

\address{Department of Physics and Astronomy, University of Calgary, Calgary, Alberta T2N 1N4, Canada}

\ead{leahy@ucalgary.ca}

\begin{abstract}
Millisecond pulsars are rapidly rotating neutron stars where general relativity plays
a strong role in the propagation of light from the neutron star to observer. 
The observed X-ray pulse shapes carry information on the mass, radius and surface shape of
the neutron star. Comparison of theoretical calculations of pulse shapes with observed pulse
shapes can give useful constraints on neutron star properties. Then comparison with 
calculated properties giving an assumed equation of state (EOS) can confirm or rule out
the assumed EOS. 
\end{abstract}

\section{Introduction}

Many neutron stars in binary systems emit X-rays from their surface during the process of
accreting matter from their companion star. There are several classes of such X-ray binaries.
One large class of X-ray binaries, the Low-Mass X-ray Binaries (LMXBs), includes  
objects in subclasses: those with optically thick X-ray emission and near Eddington-limited
X-ray luminosities (for a review see [1]) and those that are intermittent that exhibit thermonuclear powered X-ray bursts.
A second class is the bright early-discovered X-ray pulsars (such as Hercules X-1) which are rotating neutron stars with an accretion column (see [2] and references therein). However
only a very few have yielded useful mass and radius constraints, primarily due to lack of
knowledge of the emission geometry- Hercules X-1 is the exception. The class which is of
interest here is the millisecond X-ray pulsars, which were only discovered recently (e.g. see [3] for an recent list and references).
 
The answer to the question of what are the masses and radii of neutron stars has long been 
sought after. For a rewiew on this topic, see [4]. 
Measurement of the structural properties of neutron stars, in particular, of their masses 
and radii, would lead to significant restrictions on the poorly understood equation of 
state near and beyond the equilibrium density of nuclear matter. Astrophysicists have 
devised a number of methods to infer the masses and radii, since these quantities cannot
be directly measured. Masses are often reliably obtained from orbits of neutron stars in
binaries, but reliable values of radii are much harder to obtain and usually rely on a
number of assumptions combined with a model. Examples are radii obtained from the 
normalization of X-ray flux, using a distance and emissivity model. In this short paper, a
description and some results are given for the method of using the X-ray pulse shape to infer mass and radius.

\section{The pulse shape model}

Observed pulse shapes are produced by a rotating hot spot on the surface which moves in
and out of the observers line-of-sight as the star rotates (for one of the earliest models
see [5]). The pulse shape depends on the size and shape of the emission region, on the 
angular dependence of the surface emissivity (which depends on the physical process 
producing the radiation) and on the gravitational field which determines how light rays
propagate from the neutron star to the observer. For higher luminosity X-ray pulsars,
such as Hercules X-1, the emission region is a column extending above the surface of the
star ([2]), but for the lower luminosity millisecond X-ray pulsars, the emission region
is expected to be confined to the surface.

The first models for millisecond X-ray pulsars included photon propagation in the Schwarzschild metric as well as Doppler effects on frequency and intensity of light [6], 
but these gave mass and radii too small to be believable.
The extension to include the full General Relativity (GR) calculations for the neutron star 
metric and for geodesics for photons in that metric was done by [7] and [8]. 
To apply this to fitting observed pulse shapes required developing accurate enough
approximations to be able to combine the pulse shape model with a parameter-search fitting
code. These approximations include separate calculations for time-delays and Doppler effects
 for photons.
An interesting effect that was found was that the oblateness of the neutron star surface, which results from the rapid rotation, significantly affects the observed pulse shape. 
A summary of the calculations is illustrated in Fig. 1. These calculations are for a rapidly
rotating case: most millisecond pulsars are rotating between 200 Hz and 400 Hz.
The legend in Fig. 1 is as follows: Exact-full calculation using GR formalism; 
OK- oblate star surface (from GR) and Kerr metric approximation for photon propagation; 
OS- oblate star surface and Schwarzschild metric approximation; 
SK- spherical star approximation and Kerr metric; 
S+D- spherical star with Swarzschild metric. 
The fact that the differences between Kerr and exact metrics is similar to that between Schwarzschild and exact shows that frame dragging is not important for rotation
periods of 600 Hz or less. Thus we find that a suitable approximation to use for
millisecond pulsars is the OS approximation- this includes an oblate star (determined
using GR), Doppler effects, time-delays, and ray-tracing using the Schwarzschild metric.

\begin{figure}
\begin{center}
\includegraphics{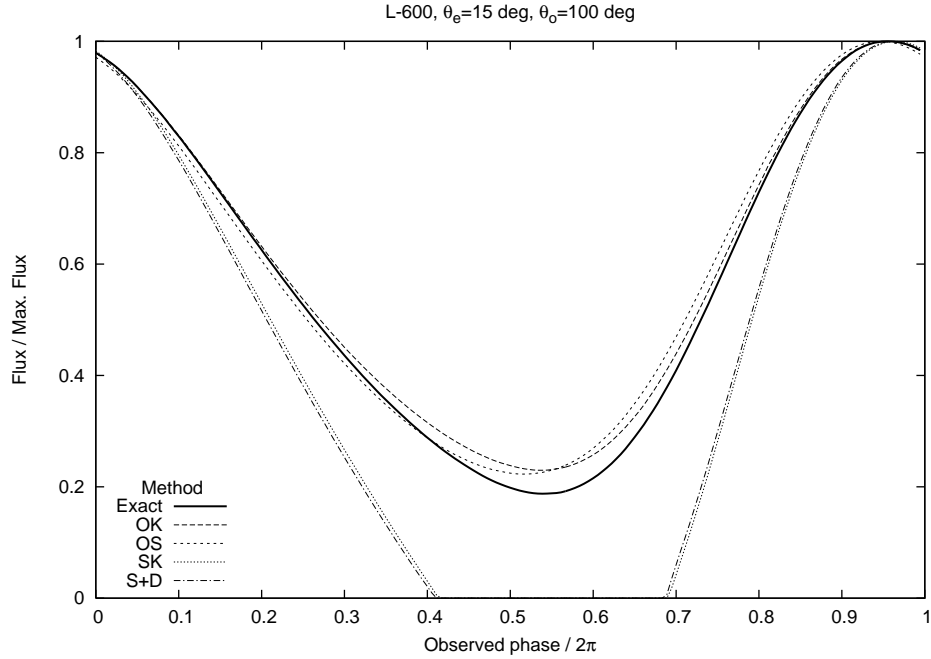}
\end{center}
\caption{\label{fig1}Light curves for a 600 Hz, 1.4 solar mass neutron star with
EOS L, emission from an angle of 15$^\circ$ from the North pole and an
observer at an inclination angle of 100$^\circ$ from the North pole. See text for legend. }
\end{figure}

\section{Comparison with data}

\begin{figure}[h]
\begin{minipage}{18pc}
\includegraphics[width=18pc]{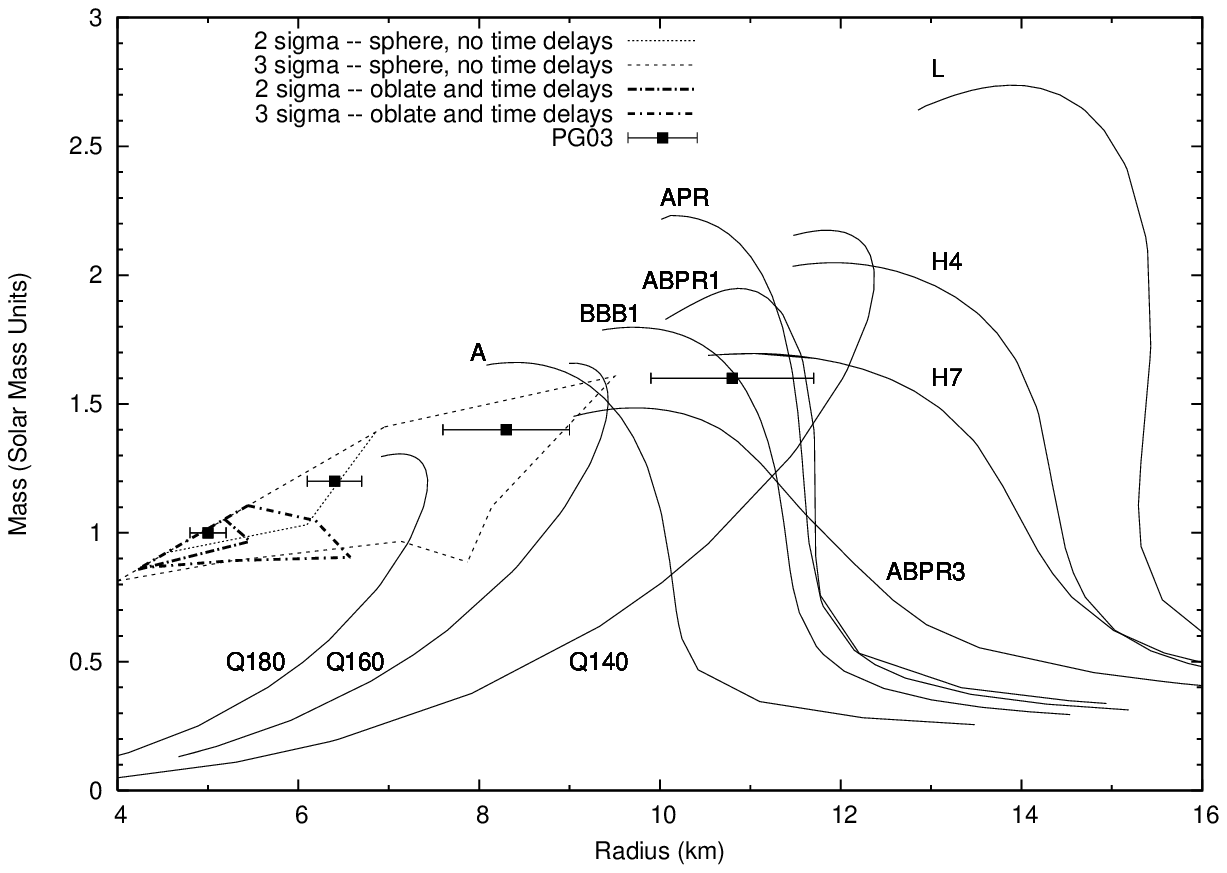}
\caption{\label{label}Mass and radius confidence contours for SAX J1808-3658 using pulse shapes in two energy bands from the 1998 outburst. }
\end{minipage}\hspace{2pc}%
\begin{minipage}{18pc}
\includegraphics[width=18pc]{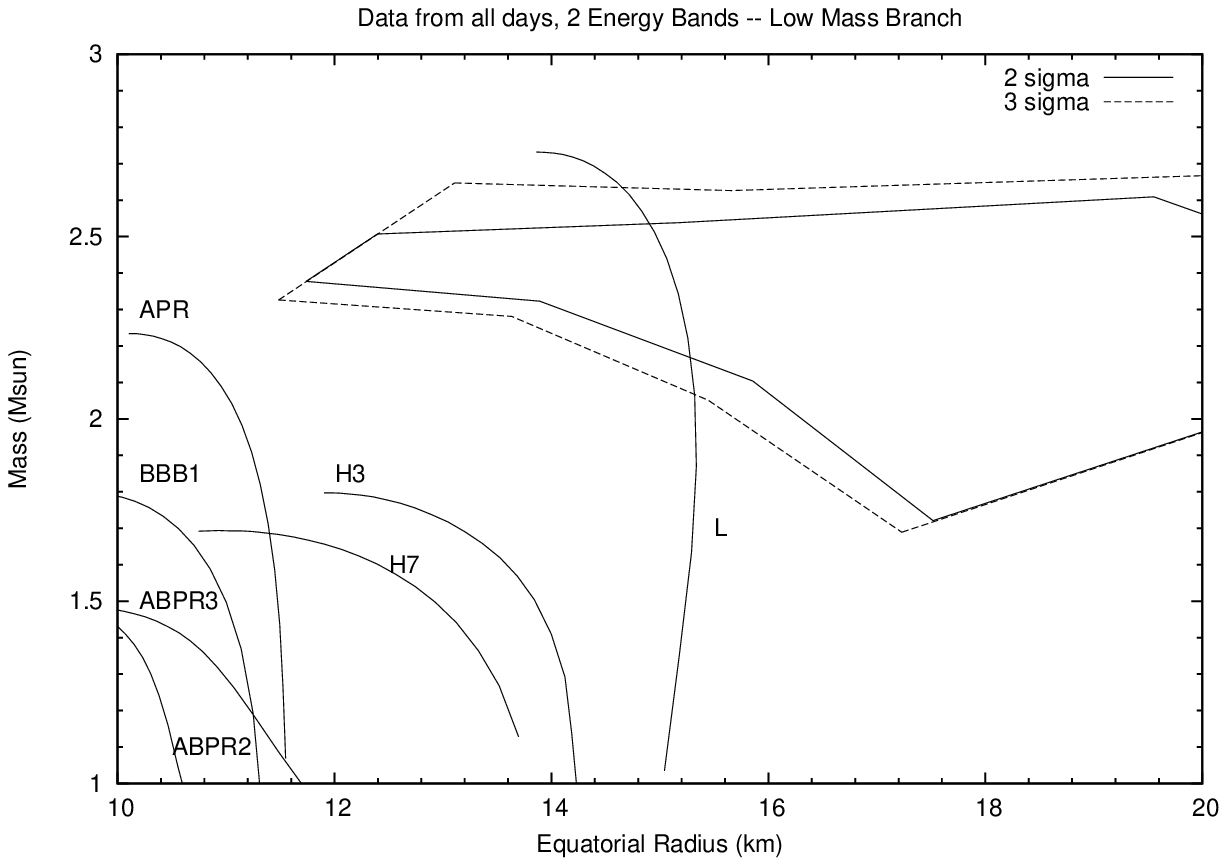}
\caption{\label{label}Best-fit mass and radius values for XTE J1814-338,using pulse shapes in two energy bands and combined data from all days. }
\end{minipage} 
\end{figure}

The OS approximation pulse shape model has been applied to Rossi X-Ray
Timing Explorer observations of the millisecond pulsar SAX J1808-3658. 
The application to the 1998 outburst pulse shape (see [9]), showed that when oblateness
and time-delays were omitted, the results of [6] were reproduced. Surprisingly
including the oblateness and time delays did not cure the problem of small inferred
mass and radius. This is illustrated in Fig. 2: the largest contour, extending from radius
of 4 to 10km and mass of 0.8 to 1.6 solar masses, is for the model omitting oblateness 
and time-delays and the smaller dot-dash contour, extending from radius
of 4 to 7km and mass of 0.8 to 1.1 solar masses, is for the full model (see [9] for description of the EOS labels).
Soon after it was realized that time variability of the pulse shape ([10]) probably played a
strong role in affecting the results.

A second millisecond X-ray pulsar, XTE J1814-338, is bright enough to have well-measured pulse shapes.
We carried out a similar analysis to that for SAX J1808-3658 [11]. For XTE J814-338, the
pulse shapes on different days (in a given energy band) were not highly variable and were
combined
into a time-averaged pulse shape. Surprisingly the pulse shape model, gave large mass
and radius for XTE J1814-338. Fig. 3 shows the 2 and 3 $\sigma$ confidence contours
which limit the mass to be within the range of 1.7 to 2.7 solar masses, 
and the radius to be within the range of 12 to 22 km (see [11] for description of the EOS 
labels).

A study has just been done [12] which includes time variability in the pulse shapes, that is,
pulse shapes from different dates are fit simultaneously with a single model. The neutron
star parameters are the same for all pulse shapes but the position of the emission region
can move on the surface between the different dates.
A single neutron star model can describe all the pulse shapes. The mass-radius constraints
are shown in Fig. 4 (see [12] for description of the EOS 
labels). The results are consistent with a number of standard EOS models.

%\begin{figure}
%\begin{center}
%\includegraphics{j1808single.eps}
%\end{center}
%\caption{\label{fig2}Mass and radius confidence contours for SAX J1808-3658 using pulse shapes in two energy bands from the 1998 outburst. Confidence contours
%corresponding to 2$\sigma$ (95.4\%) and 3$\sigma$ (99.7\%) for models assuming a spherical surface and
%no time delays are plotted as dashed curves. Confidence contours corresponding to 2$\sigma$ and
%3$\sigma$ for models assuming an oblate surface and including time delays are shown as bold dot-dashed
%curves. Best fit models computed by PG03 are shown as squares with errorbars.
%.}
%\end{figure}

%\begin{figure}
%\begin{center}
%\includegraphics{j1814all.eps}
%\end{center}
%\caption{\label{fig3}Best-fit mass and radius values for XTE J1814-338, using the combined data from all days, separated
%into two energy bands. Contours shown are for 2- and 3-s confidence levels. }
%\end{figure}

\begin{figure}
\begin{center}
\includegraphics{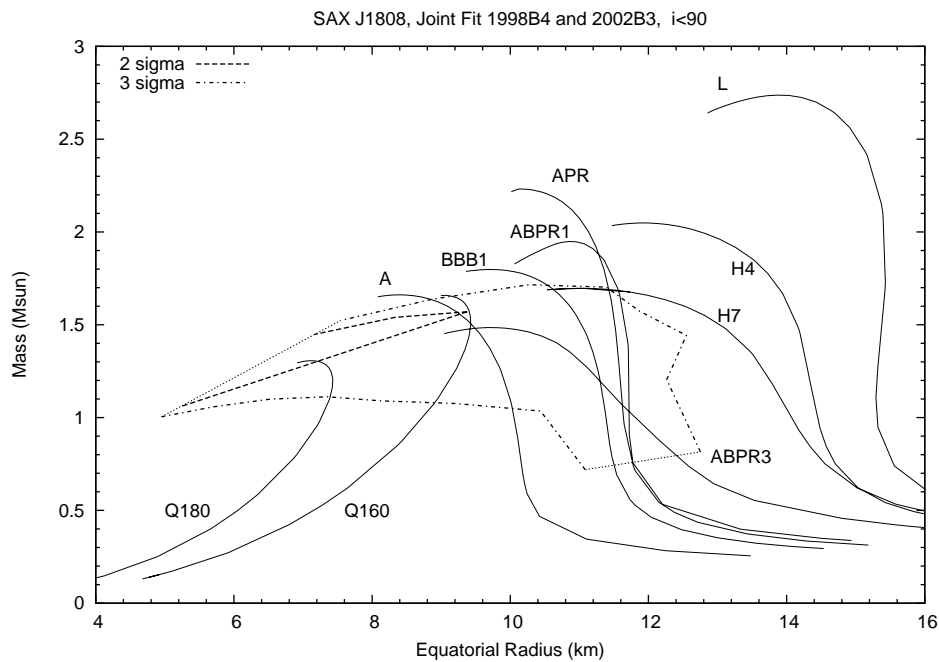}
\end{center}
\caption{\label{label}Mass and radius confidence contours for SAX J1808-3658, for joint fits to 1998B4 and 2002B3 data using the constraint inclination<90$^\circ$. Mass-radius
curves (solid curves) for stars spinning at a frequency of 401 Hz are shown for a variety of EOS.}
\end{figure}

\section{Discussion}
The latest results which take into account time-variability of pulse shapes 
in modeling pulse shapes produce satisfactory results for mass
and radius constraints. 
To reduce the size of the allowed region in the mass-radius plane will require 
better observations of pulse shapes to reduce observational errors.

%\section*{Acknowledgments}
\ack
This work supported by the Natural Sciences and Engineering Research Council of Canada.

\smallskip

\end{document}